\documentclass[12pt]{article}

\usepackage{graphics,epsf}
\usepackage{cite,a4wide}
\newcommand{\beq}{\begin{equation}}
\newcommand{\eeq}{\end{equation}}
\newcommand{\bea}{\begin{eqnarray}}
\newcommand{\eea}{\end{eqnarray}}
\renewcommand{\d}{\delta}

\renewcommand{\a}{\alpha}
\renewcommand{\ni}{\noindent}

\newcommand{\e}{\epsilon}
\newcommand{\s}{\sigma}

\newcommand{\oh}{\frac{1}{2}}

\newcommand{\non}{\nonumber}
\renewcommand{\t}{\tau}
\newcommand{\rf}[1]{(\ref{#1})}
\newcommand{\ra}{\rightarrow}
\newcommand{\pa}{\partial}

\begin{document}

\hfill January 1999

\begin{center}

\vspace{32pt}

  { \large \bf Worldsheet Fluctuations and the Heavy Quark \\
               Potential in the AdS/CFT Approach }

\end{center}

\vspace{18pt}

\begin{center}
{J. Greensite${}^{ab}$ and P. Olesen${}^c$}

\end{center}

\vspace{18pt}

\begin{tabbing}

{}~~~~~~~~~~~~~~~~~\= blah  \kill
\> ${}^a$ Physics and Astronomy Dept. San Francisco State Univ., \\
\> ~~San Francisco, CA 94117 USA.  E-mail: {\tt greensit@stars.sfsu.edu} \\
\\
\> ${}^b$ Theoretical Physics Group, Mail Stop 50A-5101, \\
\> ~~Lawrence Berkeley National Laboratory, Berkeley, CA 94720 USA. \\ 
\\
\> ${}^c$ The Niels Bohr Institute, Blegdamsvej 17, \\
\> ~~DK-2100 Copenhagen \O, Denmark.  E-mail: {\tt polesen@nbi.dk}

\end{tabbing}

\vspace{18pt}

\begin{center}

{\bf Abstract}

\end{center}

\bigskip

   We consider contributions to the heavy quark potential, in the
AdS/CFT approach to SU(N) gauge theory, which arise from first order
fluctuations of the associated worldsheet in anti-deSitter space.
The gaussian fluctuations occur around a classical worldsheet configuration
resembling an infinite square well, with the bottom of
the well lying at the AdS horizon.  The eigenvalues of the 
corresponding Laplacian operators can be shown numerically to be very 
close to those in flat space. We find that two of the transverse world sheet 
fields become massive, which may have implications for the existence
of a L{\"u}scher term in the heavy quark potential. It is also suggested
that these massive degrees of freedom may relate to extrinsic curvature 
in an effective $D=4$ string theory.

\vfill

\newpage
\section{Introduction}

Maldacena's conjecture \cite{Mal}, relating the large $N$ expansion
of conformal fields to string theory in a non-trivial geometry, has
led to the hope that non-perturbative features of large $N$ theories may be
understood. Witten's extension \cite{Witten} of this conjecture 
to non-supersymmetric gauge theories, such as large $N$ QCD in four dimensions,
provides a new and elegant approach to the study of gauge theory at strong
couplings.

In Witten's approach the heavy quark potential has a linear behaviour 
\cite{Witten,Mal1,Brand,Rey}. In this approach the temperature $T$ in 
the higher dimensional theory acts as an ultraviolet cutoff, and the strong
coupling $g_{YM}^2N$ is the bare coupling at the scale $T$. The problem is,
of course, how to extend this to lower coupling, and whether
one encounters a phase transition on the way, as discussed in \cite{Gross}.

In the approach of refs. \cite{Witten,Mal1,Brand,Rey} the interquark
potential has been extracted at the saddle point. In the present paper
we extend this by including fluctuations of the world sheet to first order.
The present paper was initiated as a sequel to a previous letter \cite{Us}, 
where we have called attention to two features of 
strong-coupling, planar $QCD_3$ in the saddle point approximation, which 
do not entirely agree with 
expectations based on lattice QCD.  First, there is the fact that the 
glueball mass is essentially independent of string tension in 
the strong-coupling supergravity
calculation \cite{Hirosi}, and goes to a finite constant in the $\s\ra \infty$
limit.  This is quite different from the behavior in strong-coupling
lattice gauge theory, where a glueball is understood as a closed loop
of electric flux whose mass tends to infinity in the infinite tension
limit, and it suggests that truly different physical mechanisms may underlie
the mass gap in the two cases. The second point concerns the existence of a 
universal L{\"u}scher term of the form $-c/L$ in the interquark potential. 
Here $c$ is a numerical, coupling independent, constant.  Recent lattice Monte 
Carlo simulations \cite{Teper} indicate the presence of such a term in 
$QCD_3$, with a value of $c$ consistent with that of a bosonic string, 
although there is a caveat that $-c/L$ represents a quite small 
correction to the dominating linear potential, and the magnitude of $c$ 
is not yet well determined numerically. Following the approach of refs.\ 
\cite{Witten,Mal1,Brand,Rey}, we have found that the interquark potential 
extracted from the saddle point action of a classical worldsheet in
$AdS_5 \times S_5$, has no L{\"u}scher term at all, which seems to
contradict the existing trend in the Monte Carlo data.

    It is quite possible, however, that the L{\"u}scher term arises beyond 
the classical worldsheet approximation, when quantum fluctuations of the 
worldsheet in $AdS_5 \times S_5$ are taken into account 
\cite{Hirosi1,KT,Sonnen}. This question is the main motivation for the work 
reported in the present paper. 

In Section 2 we study the background field in the saddle point for large 
interquark distances. It turns out that the radial AdS coordinate 
$U$  \cite{Mal} of the string worldsheet is situated at the horizon, 
except for a small interval in parameter space near the end points 
$\sigma=\pm L/2$, where $U$ is forced to shoot
up to infinity. In Section 3 we introduce Kruskal-like coordinates, and 
discuss the near-flatness of this metric at the horizon, in the
$g^2_{YM}N \ra \infty$ limit. The eigenvalues 
and eigenfunctions for the relevant Laplacians are then shown to be 
essentially the same as in the completely flat case, 
with the contour of the classical worldsheet bringing the problem into
the form of an infinite square well.

In Section 4 we discuss the expansion of the action to the first
non-trivial order. It is found that two of the transverse worldsheet
coordinates become massive, and do not contribute to the L{\"u}scher
term.  We argue that, due to the vanishing curvature in the
$g^2_{YM}N \ra \infty$ limit, the fermion and ghost contributions will
have essentially flat-space contributions to $-c/L$, although we do not
claim to show this explicitly.  

   If the fermionic and ghost contributions are in fact similar to
flat space, then a possible consequence is that the L{\"u}scher term has a
sign opposite to the one extracted from the Monte Carlo data.
The L{\"u}scher term is essentially the Casimir energy of a
string with fixed ends.  For a superstring in flat space with Ramond boundary
conditions, there is an exact cancellation of bosonic and fermionic
contributions to the Casimir energy, and the  L{\"u}scher term vanishes.
In the case of Neveu-Schwartz boundary conditions, bosonic and fermionic
zero-point energies contribute with the same sign, and this is what leads
to a tachyonic state for a free string.  However, when the GSO projection 
is taken into account and the tachyonic state is removed, we again have
a massless ground state and a vanishing L{\"u}scher term.  The relation
between having a tachyonic ground state for the free string, and
an attractive potential from the L{\"u}scher term, is discussed in
ref.\ \cite{poul}.  

   Since the string worldsheet, in the AdS/CFT approach, is intended to
describe the dynamics of the QCD string between massive quarks, then 
presumably the appropriate boundary conditions are Neveu-Schwartz, with
a GSO projection removing the lowest state.
Taking account of the mass term for two of the transverse degrees
of freedom, and assuming (as curvature becomes negigible in the
$g^2_{YM}N \ra \infty$ limit) that the rest of the 
degrees of freedom contribute to the Casimir energy as they do in flat space,  
the result is a ``wrong-sign'' L{\"u}scher term.  We must emphasize,
however, that this is a very tentative conclusion, and one which does not take
the Ramond-Ramond background into consideration.

  So far these results refer to QCD in three dimensions. Section 5 contains
a brief discussion of the four dimensional case.
Finally, in section 6, we suggest that in four dimensions the massive 
world sheet fields may relate to extrinsic curvature terms in an effective
4D string theory. In Section 7 we conclude. It is noted that if
our tentative result for the sign of the L\"uscher term is correct,
it implies that the strong coupling supergravity approach to QCD does not 
correspond to QCD defined on a lattice.

\setcounter{equation}{0}
\section{The saddle point field for large interquark distances}   

   As explained in ref.\ \cite{Witten,Brand}, spatial Wilson loops in 
D=3 planar Yang-Mills theory are computed, in the supergravity approach,
from the dynamics of worldsheets in the near-extremal background
metric 
\begin{equation}
ds^2=\alpha'\left\{\frac{U^2}{R^2}\left((1-U_T^4/U^4)dt^2+
  \sum_i dx_i^2\right)+\frac{R^2}{U^2}\frac{dU^2}
{1-U_T^4/U^4}+R^2d\Omega_5^2\right\}.
\label{metric}
\end{equation}
The boundary of the worldsheet is a rectangle in the $x_1-x_2$ plane
at $U=\infty$, whose interior, specified by $x_1=\sigma,~x_2=\tau$ with 
$|\sigma|\leq {L\over 2}$, and $|\tau|\leq {Y\over 2}$, 
parametrizes the worldsheet
of a $L\times Y$ Wilson loop with $Y \gg L$.  The classical worldsheet,
in the $Y \ra \infty$ limit, is given by $x_1(\s,\t)=\s,~x_2(\s,\t)=\t$,
and $U(\s)$ determined implicitly from
\begin{equation}
\frac{L}{2}-\sigma=\frac{R^2}{U_0}\int_{U/U_0}^\infty \frac{dy}{\sqrt{(y^4-1)
(y^4-1+\epsilon)}}
\end{equation}
with 
\beq
U_0=U(\s=0),~~~~ \e=1-U_T^4/U_0^4,~~~~ R^2 = \sqrt{4\pi g^2_{YM}N}, 
~~~~ U_T = R^2 b
\eeq

\begin{figure}[ht]
\centerline{\scalebox{.5}{\rotatebox{270}{\includegraphics{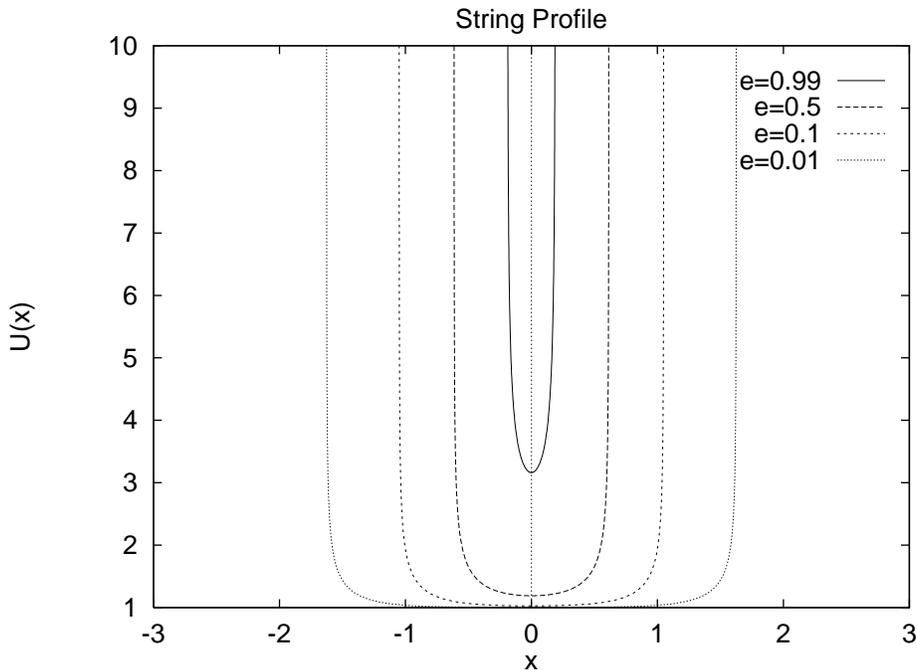}}}}
\caption{String contours $U(x)$ for various $\e = 1-(U_T^4/U_0^4)$,
in units of $U_T=R^2b$.  The asymptotes of each curve lie at $x=\pm L/2$.
Note the approach to the horizon (here at $U=1$), as $\e \ra 0$.}
\label{fig1}
\end{figure}

   The metric (\ref{metric}) is relevant for the 
calculation of the boson and fermion contributions to the action. In general,
since the background field $U=U(\sigma)$ is a non-trivial function
of $\sigma$, one cannot expect that world 
sheet supersymmetry is preserved in the presence of this
background field.  On the other hand, a graph of $U(\s)$ in the range 
$\s \in [-{L\over 2},{L\over 2}]$ looks very much like an infinite 
square well at large $L$, as seen in Fig.\ \ref{fig1}.  Starting
at $U(-{L\over 2})=\infty$, $U(\s)$ drops precipitously to $U(\s)\approx U_0
\approx U_T$, remaining almost constant in a range 
$[{L\over 2}+d,{L\over 2}-d]$ 
where $d\ll L$, and then shoots back up to $U=\infty$ at $\s={L\over 2}$.  
The fact that the classical worldsheet coordinate $U(\s)$ is nearly
constant for most of the range of $\s$ is, of course, very relevant for a 
saddlepoint calculation, where we include the effect of gaussian fluctuations 
around the classical worldsheet.  

  We will need expressions for $U(\s)$ both near and away from 
$\s = \pm {L\over 2}$.  Denoting  $y(\s) = U(\s)/U_0$, we have
\begin{equation}
dy/d\sigma=b\sqrt{(y^4-1)(y^4-1+\epsilon)}
\label{1}
\end{equation}
where $\epsilon=1-U_T^4/U_0^4$ was found \cite{Us} to be related to the 
interquark distance $L$ by
\begin{equation}
\epsilon\approx e^{-2bL}.
\end{equation}
Away from the endpoints at $\s=\pm {L\over 2}$ make the trial expansion
\begin{equation}
y(\sigma)\approx 1+\delta (\sigma),~{\rm with}~|\delta(\sigma)|\ll 1. 
\end{equation}
and then linearize eq.(\ref{1}), 
\begin{equation}
d\delta/d\sigma\approx 2b\sqrt{\delta (4\delta+\epsilon)},
\end{equation}
which is valid as long as $\delta$ stays small. Integrating we obtain
\begin{equation}
\ln (2\sqrt{\delta}+\sqrt{4\delta+\epsilon})=2b\sigma+\ln\sqrt{\epsilon},
\end{equation}
where we used the boundary condition that $y=1$,  and hence $\delta=0$, for
$\sigma=0$. Solving this equation for $\delta$, we get
\begin{equation} 
y(\sigma)\approx 1-\exp (-2bL)/8+[\exp(-4b(L/2+\sigma))+
\exp(-4b(L/2-\sigma))]/16. 
\label{solution}
\end{equation}
Thus, for $|\sigma|<L/2$ the corrections to $y=1$ are exponentially small,
and $U(\s)\approx U_0$ is essentially constant. 

\begin{figure}[tb]
\hspace*{2.5cm}
\epsfbox{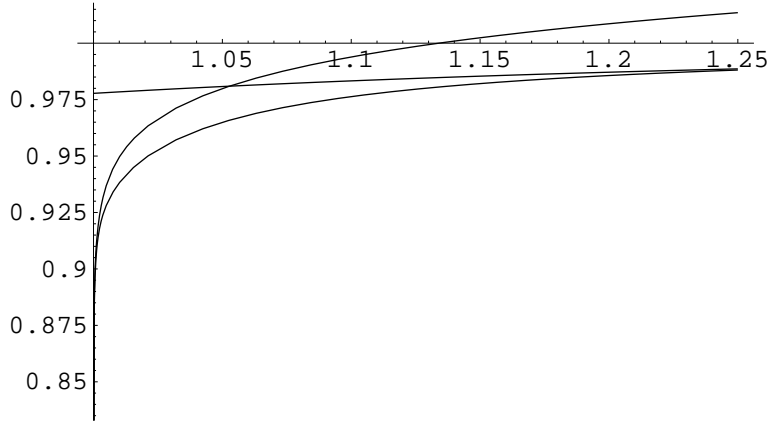}
\caption{A $\sigma/(L/2)$ versus $U/U_0$ plot of the exact solution for 
$bL=30$, compared to the two 
asymptotic solutions. The solution valid for $\sigma\approx L/2$ approaches
the exact solution near $y=1.25$, whereas the $|\sigma| <L/2$ solution
starts to deviate from the exact solution near $\sigma\approx 0.925 (L/2)$.}
\label{figure1}
\end{figure}

   For $|\sigma|\rightarrow {L\over 2}$ 
this analysis breaks down, since $\d$ is
not small. From the relation
\begin{equation}
\frac{L}{2}-\sigma=\frac{R^2}{U_0}\int_{U/U_0}^\infty \frac{dy}{\sqrt{(y^4-1)
(y^4-1+\epsilon)}}\approx\frac{R^2U_0^2}{3U^3}\approx\frac{1}{3by^3},
\end{equation}
using $U_0\approx U_T=R^2b$, we see that
\begin{equation}
U\approx\frac{R^2b}{(3b(L/2-\sigma))^{1/3}},~{\rm for}~
\sigma\rightarrow L/2
\label{end}
\end{equation}
in the neighbourhood of $|\sigma|\rightarrow {L\over 2}$.  A plot of the
exact solution for $y(\s)$ at $bL=30$, and the two asymptotic
solutions \rf{solution} and \rf{end}, is shown in Fig.\ \ref{figure1}.

\begin{figure}[tb]
\hspace*{2.5cm}
\epsfbox{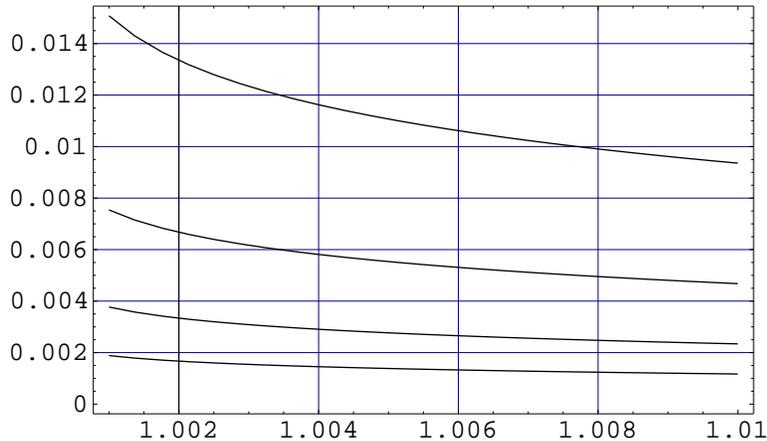}
\caption{A plot of $1-\sigma/(L/2)$ as a function of $y=U/U_0$. The
curves from the top to the bottom correspond to $bL$=200, 400, 800, and
1600, respectively. If we fix the upper limit on $d$ by requiring that
$y$ should only deviate from 1 by $10^{-3}$, corresponding
to the left side of the figure, we see that $1-\sigma/(L/2)$ decreases
like $1/L$ to a high accuracy, because in going from the top to the
bottom at $y=1.001$, the distance between the successive curves decreases by
a factor two. }
\label{figuren}
\end{figure}

   According to eq. \rf{solution} and Fig.\ \ref{fig1}, the classical
solution for $U(\s)$ is almost
constant in some interval $[-{L\over 2}+d,{L\over 2}-d]$.
To estimate $d$, we can first ask for the value close to
$\sigma={L\over 2}$ where the asymptotic solutions \rf{solution} and \rf{end}
are equal. This happens for
\begin{equation}
\frac{\sigma}{L/2}\approx 1-\frac{.63}{bL}.
\label{estimate}
\end{equation}
A more stringent criterion, arrived at numerically, is to ask where
$y(\s)$ deviates from $y=1$, at large $L$, by more than $10^{-3}$. 
With this criterion for $d$, we find that $d<1.5/b$, approximately, 
obtained from the solutions for $y(\s)$ at various $L$ shown in 
Fig.\ \ref{figuren}. 

\setcounter{equation}{0}
\section{Eigenvalues of Laplacians in the AdS background}

  We would like to make an expansion around the saddle point. In order to do
this, it is convenient to use different variables than $U$ and $t$, because of
the singular form of the metric (\ref{metric}). We therefore introduce the 
Kruskal-like coordinates for $U>U_T$
\begin{eqnarray} 
&&T=\frac{\sqrt{2}~R^2}{U_T}~e^{-\pi/4}e^{\tan^{-1}(U/U_T)}
\sqrt{\frac{U-U_T}{U+U_T}}\cos\left(\frac{2U_T t}
{R^2}\right),\nonumber \\
&&Z=\frac{\sqrt{2}~R^2}{U_T}~e^{-\pi/4}e^{\tan^{-1}(U/U_T)}
\sqrt{\frac{U-U_T}{U+U_T}}\sin\left(\frac{2U_T t}
{R^2}\right).
\label{newvariables}
\end{eqnarray}
These expressions are valid in Euclidean space, and in Minkowski space
the sine and cosine are replaced by the hyperbolic sine and cosine,
respectively. The time variable $t$ is periodically identified by
$t\rightarrow t+\pi/b$, with $U_T=b R^2$.
In these coordinates
\begin{eqnarray}
ds^2&=&\alpha'\left\{\frac{(U^2+U_T^2)(U+U_T)^2e^{\pi/2}}{8 R^2U^2}
e^{-2\tan^{-1}(U/U_T)}\left(dT^2+dZ^2\right)\right.\nonumber \\
&+&\left.\frac{U^2}{R^2}\sum_i dx_i^2+R^2d\Omega_5^2\right\},
\label{newmetric}
\end{eqnarray}
so that the metric is now symmetric in terms of the new variables
$Z$ and $T$. As usual with the (Euclidean) Kruskal metric, $U$ should be
considered as a function of $T^2+Z^2$ through the equation ($U>U_T$)
\begin{equation}
Z^2+T^2=\frac{2R^4 e^{-\pi/2}}{U_T^2}e^{2\tan^{-1}(U/U_T)}\frac{U-U_T}{U+U_T}.
\end{equation}
It should be noticed that the metric (\ref{newmetric}) is flat up to
exponentially small terms, except at the end points $\sigma=\pm L/2$.

   The saddlepoint contribution to the spatial Wilson loop is given by
simply evaluating the Nambu action of the classical worldsheet in this
metric \cite{Witten,Brand}, and is found to be
\beq
       S_{cl} =  {U_T^2 \over 2\pi R^2} Y L 
\eeq
We are interested now in the contribution from gaussian fluctuations around
the saddlepoint, which involve the bosonic, fermionic, and ghost degrees
of freedom, in the limit of very large $R$.    

   In the $R\ra \infty$ limit the curvature of the 5-sphere (as well
as the curvature of AdS space) vanishes,
and the contribution of each degree of freedom associated with the 5-sphere
is identical to the corresponding flat-space value, i.e.\ $-\pi Y/24 L$.
Likewise, fluctuations around the classical worldsheet in AdS space
in the neighborhood of the horizon, i.e.\ 
$\s \in [-{L\over 2}+d, {L\over 2}-d]$, are essentially fluctuations
in flat space, and the relevant differential operators are either the
flat-space 2D Laplacian, or, as we shall see in the next section, this
operator plus a mass term. 
  Thus, for example, the eigenstates $\psi(\s,\t)$ of
\beq
      \pa^2_M \equiv \pa_a G_{MM}[U_{cl}(\s)] \pa^a
\eeq
will be identical to eigenstates of the flat-space 2D Laplacian, i.e.
\beq
        \psi(\s,\t) \propto \sin[\a (\s+ c)] e^{i\omega \t}
\label{propto}
\eeq 
away from the $\s=\pm {L\over 2}$ endpoints.  The eigenvalue spectrum is
determined by the boundary conditions $\psi(\s,\t)=0$ at $\s=\pm {L\over 2}$ 
(meaning that fluctuations vanish at the Wilson loop perimeter).
In flat space these conditions yield the usual result that
\beq
      \a^{flat}_n = {n\pi \over L} ~~, ~~~~~~ c = L/2
\label{flad}
\eeq
In AdS space the values for $\a$ are slightly different, owing to the
fact that eq. \rf{propto} breaks down for ${L\over 2} - |\s|<d$.  Very close
to the endpoints, the operator $\pa^2_M$ becomes $\pa_a U^2(\s) \pa_a$.
We solve for the eigenfunctions in this region by making
separation ansatz $\psi (\tau,\sigma)=\Theta (\tau)S(\sigma)$, and find for 
the eigenvalue equation $\partial_i(U^2\partial_i)\psi=\Lambda\psi$ near 
the end points
\begin{eqnarray}
&&\pa_\s^2 S+\frac{2}{3(L/2-\sigma)}\pa_\s S-\tilde\Lambda 
(L/2-\sigma)^{2/3}S-
\lambda S=0,~
{\rm where}~\tilde\Lambda=\frac{(3b)^{2/3}}{R^4b^2}\Lambda.
\nonumber \\
&&\pa^2_\tau \Theta= -\lambda \Theta.
\label{separation}
\end{eqnarray}
Here $\lambda$ is a separation constant. The equation for $\Theta$ is the 
same as for the $\pa^2$ operator, whereas for the function $S$ in the
neighborhood of the endpoints there are two 
solutions, namely one for which $S$ vanishes, in $\s \ra {L\over 2}$ limit, 
as
\begin{equation}
S\approx{\rm const.}~(L/2-\sigma)^{5/3}
\label{endpoint}
\end{equation}
and one where $S$ goes to a non-zero constant for 
$\sigma\rightarrow {L\over 2}$.
The solution vanishing at the endpoints is the one which is relevant
for worldsheet fluctuations.  Away from the endpoints, $\psi(\s,\t)$ has
the harmonic form shown in eq. \rf{propto}.
The ``end point solution'' (\ref{endpoint}) vanishes more rapidly
than the sine function near $\s=\pm {L\over 2}$,
which is due to the fact that in eq.
(\ref{separation}) the first derivative $S'$ is multiplied by a large
factor, and hence is forced to be small. 

   We can now make a rough estimate of how the eigenvalues of $\pa^2_M$ 
compare to those of the flat-space operator, based on the fact that 
$\psi(\s,\t)$ falls much more rapidly to zero,
near the endpoints at $\s=\pm {L\over 2}$, than the sine function. 
This allows us to approximate 
$\psi(\s,\t)$ as a harmonic function in the range
$[-{L\over 2}+d,{L\over 2}-d]$, and equal to zero outside this range.
Then
\bea
     {|\a_n - \a^{flat}_n| \over \a^{flat}_n} &\sim& 
     { {n \pi \over L-d} - {n \pi \over L} \over {n \pi \over L} }
\non \\
            &\sim& O\Bigl[{d \over L}\Bigr]
\non \\
            &\sim& O\Bigl[{1 \over bL}\Bigr]
\eea
Since the flat-space eigenvalues for the massless Laplacian lead to a 
L\"uscher term of $O(1/L)$, these
small deviations can only lead to a further correction, in the AdS case, 
of still higher order in $1/L$. For the massive Laplacian the situation
is, however, different, as we shall see in the next section.  

   A similar observation presumably applies to the fermionic and ghost 
degrees of freedom The associated differential operators in $\s,~\t$ 
again only deviate from the corresponding flat-space case in a 
region near the endpoints, where the derivatives are multiplied by a 
factor of $U(\s)$; this region is a very small fraction (of
order $1/L$) of the full interval.  Eigenmodes of these operators will
have to be nearly constant in the ``shootup'' region near the endpoints,
where $U(\s) \ra \infty$.  However, as in the case of the bosonic modes, this 
slight modification of the eigenmodes will only affect the values of the 
determinants at higher orders in $1/L$.  

\setcounter{equation}{0}
\section{The bosonic action and the necessity of massive fields}

We want now to study the bosonic action, keeping only quadratic terms in the 8
transverse variables ($Z,T,x_3,...$). We start from the partition function
\beq
{\cal Z}=\int {\cal D}X \sqrt{G}\exp \left(\frac{1}{2\pi\alpha'}\int d^2\sigma
\sqrt{\det G_{MN}(X(\sigma))\pa_a X^M\pa_bX^N}\right),
\label{p}
\eeq
where we integrate over the 10 variables $X^M$, and insert a factor $\sqrt{G}$
in order to have a measure which is invariant with respect to changes of the
coordinates entering the $AdS_5\times S_5$. We also want to
choose a gauge where $\sigma,\tau$ are identified with $x_1,x_2$.
The measure factor in (\ref{p}) can then be exponentiated in the form
\beq
\sqrt{G}=\exp \left(\Lambda^2\int d^2\sigma\sqrt{h}\ln G\right).
\eeq
Here $h$ is the measure associated with the world sheet variables, so
$\sqrt{h}=\alpha' U^2/R^2$, and $\Lambda$ is a ultraviolet cutoff.
This form of the exponentiation is reparametrization invariant.
Because of the absence of a $1/2\pi\alpha'$ factor in the exponentiated
version of $\sqrt{G}$, this factor will only contribute to terms of order
$\alpha'$ in the effective action. We shall not consider this order, and we 
therefore ignore the $\sqrt{G}$ contribution in the following.

Now, if we expand the action, keeping only second order terms,
we get
\begin{equation}
S\approx (1/2\pi)\int d^2\sigma\left\{ U^2/R^2+(1/2)\left[(U^2_T/R^2)\left(
(\partial_iZ)^2
+(\partial_iT)^2+(\partial_i x_3)^2\right)+
R^2(\partial_iy^M)^2\right]\right\},
\label{ng} 
\end{equation}
where the $y$'s refer to the 5-sphere, and where we took $x_1,x_2$ to be
longitudinal. Of course, it is important to keep {\sl all} second order 
terms. To this end, we need to notice that the $U^2$ in the first term is 
given as a
function of $Z,T$. Exactly at the
horizon $Z=T=0$, and $Z^2+T^2$ therefore represent the small, second order
deviations of the radial variable from its value at the horizon,
\begin{equation}
U\approx U_T+(U_T^3/R^4)(Z^2+T^2).
\end{equation}
Inserting in eq. (\ref{ng}), we find to 2nd order in the fluctuations
\begin{eqnarray}
S&\approx& (1/2\pi)\int d^2\sigma( U_T^2/R^2+
(U_T^2/2R^2)[(\partial_iZ)^2
+(\partial_iT)^2+(4U_T^2/R^4)(Z^2+T^2)]\nonumber \\
&& +(U_T^2/2R^2)(\partial_i x_3)^2)
+(R^2/2)(\partial_iy^M)^2),
\label{ng2} 
\end{eqnarray}
which shows that the fields $Z,T$ have mass terms with coefficients 
$4U_T^2/R^4=4b^2$. Thus two bosonic degrees of freedom, originally
associated with the $U,~t$ coordinates, have become massive, and it
is not hard to see why such a ``potential'' term must exist: The boundary
of the worldsheet lies at $U=\infty$, yet the preferred position of
the string, as $L \ra \infty$, lies at the black hole horizon $Z=T=0$. 
The first term in the integral gives the leading contribution 
$(U_T^2/2\pi R^2)YL$, corresponding to a 3D string tension 
\beq
      {\cal T}_3  = {U_T^2 \over 2\pi R^2} = {R^2 b^2 \over 2\pi}
\label{lowest}
\eeq
derived in refs.\ \cite{Witten,Brand,Rey}.

The Gaussian integral over $Z,T$ can be performed, e.g. by use of
analytic regularization \cite{sungkil} ($\ln x=\pa x^\beta /\pa \beta$ for 
$\beta\rightarrow 0$). Since $Y\rightarrow\infty$, the sum over the
``time-eigenvalues'' can be replaced by an integral, which can be
performed to give
\begin{equation}
{\rm tr~ln}(-\nabla^2+4b^2)=-(Y/\sqrt{4\pi})(\partial/\partial\beta)
(\Gamma (\beta-1/2)/
\Gamma (\beta))\sum_{n=1}^\infty ((n\pi/L)^2+(4b^2))^{-\beta+1/2}
\end{equation}
with $\beta\rightarrow 0$. 
The sum over $n$ can be carried out and the limit $\beta\rightarrow 0$
can be performed to give \cite{sungkil}
\begin{equation}
{\rm tr~ln}(-\nabla^2+4b^2)=-\frac{YL}{\pi}b^2(-1+\ln (4b^2/\mu^2))-
Yb\left[1+\frac{2}{\pi}\sum_{n=1}^\infty\frac{1}{n}K_1(4nbL)\right].
\label{dk}
\end{equation}
Here $\mu$ is an ultraviolet cutoff, which occurs in the heat kernel method,
which gives cutoff dependent terms proportional to $\mu^2$ and $b^2\ln\mu^2$.
The $\mu^2$ terms are present for all fields, and if we add the
fermions they cancel completely. The logarithmic terms only occur for
the massive fields (see e.g. \cite{sungkil}), and they combine with the
$b^2\ln (4b^2)$ term to give the result exhibited in (\ref{dk}).
 
Using the asymptotic expansion of the Bessel function valid for large
$L$, we get
\begin{equation}
{\rm tr~ln}(-\nabla^2+4b^2)\approx-\frac{YL}{\pi}b^2(-1+\ln (4b^2/\mu^2))
-Yb\left[1+\frac{1}{\sqrt{2\pi bL}}e^{-4bL}+...\right].
\label{masss}
\end{equation} 
This can be compared to the massless case,
\beq
{\rm tr~ln}(-\nabla^2)=-\pi Y/12L.
\eeq
It can be shown that this result follows by rewriting the sum over
Bessel functions in eq. (\ref{masss}), by use of the following relation
\begin{eqnarray}
&&\sum_{n=1}^\infty K_1(nz)/n=\pi^2/6z+(1/4){\bf C}z+(1/8)z
\ln (z/(4\pi)^2)-z/16+\pi/4\nonumber \\
&+&\pi\sum_{l=1}^\infty\left(\sqrt{1+4l^2\pi^2/z^2}
-2l\pi/z-z/4l\pi\right),
\end{eqnarray}
where $\bf C$ is Euler's constant, and taking the limit $b\rightarrow 0$. 
The first term on the right hand side gives the desired result for $b
\rightarrow 0$ if we take $z=4bL$. For $bL$ large, the above expression is not
useful, and the asymptotic expansion of the Bessel functions should then be 
used.

   We have stressed, in the previous section, that curvature in
$AdS_5 \times S_5$ tends to zero in the $R \ra \infty$ limit, and string
fluctuations in the neighborhood of the horizon are essentially fluctuations
in a flat-space metric.  That being the case, how can we find a mass term
in eq.\ \rf{masss} of $O(b^2)$, which is finite in the $R \ra \infty$
limit?  At first sight, this seems a violation of the principle of 
equivalence.  To understand what is going on, we first note that the metric
coefficients in eq.\ \rf{newmetric} are all of order $R^2$ near the
horizon.  The integration in \rf{ng2} runs from $-L/2$ to $+L/2$,
but in fact the proper time along the horizon is of order $RL$.  If we 
make a trivial change
of variables, simply rescaling all coordinates (and parameters $\s,\t$)
by a factor of $R$ so metric coefficients are all $O(1)$ near the horizon, 
then the contribution to the action from the region along the horizon
is approximately
\begin{eqnarray}
S&\approx& (b^2/2\pi)\int^{+RY/2}_{-RY/2} d\t \int^{+RL/2}_{-RL/2} d\s
 \Bigl( 1 + \oh [(\partial_iZ)^2
+(\partial_iT)^2+(4b^2/R^2)(Z^2+T^2)] 
\nonumber \\
&& +\oh(\partial_i x_3)^2)
+(1/2b^2)(\partial_iy^M)^2 \Bigr),
\label{ng3} 
\end{eqnarray}
Here the mass term evidently tends to zero as $R\ra \infty$, as one
would expect from the equivalence principle.  But this decrease is 
precisely compensated by the growth of the worldsheet along the
horizon (as seen in the limits of integration) as $R$ increases.
The end result of a gaussian integration is, of course, identical to
eq. \rf{masss}; one finds a finite, $R$-independent mass term in the 
trace log.

  For the bosonic part we thus have two massive and six
massless degrees of freedom. The contribution from the bosonic part of
the string to the potential is thus
\beq
{\rm Potential~from~bosons}=\frac{R^2b^2}{2\pi}\left(1-\frac{2}{R^2}
\ln \frac{4b^2}{e\mu^2}\right)L-\frac{\pi}{4L}.
\eeq
We see that bosonic contributions are responsible for a logarithmic
correction to the lowest order result for the string tension \rf{lowest},
i.e.
\beq
{\cal T}_3=\frac{R^2b^2}{2\pi}\left(1-\frac{2}{R^2}
\ln \frac{4b^2}{e\mu^2}\right).
\eeq
As $g^2_{YM}N\ra \infty$, the curvature of AdS space tends to zero.
If the contributions from the fermions and ghosts can really be 
obtained in the flat space limit near the horizon, as argued in the
last section, then the resulting
L\"uscher term would be the same as if the calculation were done in
flat space, with the contribution $-\pi/12L$ from two transverse bosonic modes
removed.  With either Ramond boundary conditions, or Neveu-Schwartz
boundary conditions with the tachyon projected out, the result is a
L\"uscher term $+\pi/12L$.  This term has the opposite sign relative 
to what is seen in lattice calculations. However, the fermions in the
full AdS background really need to be investigated further,
before this can be considered as a safe conclusion. 

\setcounter{equation}{0}
\section{The potential in four dimensions}

Let us consider the relevant metric \cite{Brand} near the 
horizon $U\approx U_T$,
\begin{equation}
\frac{ds^2}{\alpha'}\approx
\frac{R^{3/2}}{3U_T^{1/2}}\frac{dU^2}{U-U_T}+
\frac{3U_T^{1/2}}{R^{3/2}}(U-U_T)dt^2\equiv dr^2+r^2d\theta^2\equiv dX^2+dY^2,
\end{equation}
with $X=r\cos\theta,~Y=r\sin\theta$ ($X,Y$ thus correspond to the coordinates
previously denoted by $T,Z$ in the three dimensional case). Here we left out 
the four-sphere as well
as the four $x$-coordinates, since these are not important for the following. 
Instead of finding the full Kruskal
coordinates, we only look at the local ones near the horizon,
\begin{equation}
dr=\frac{R^{3/4}}{\sqrt{3}U_T^{1/4}}\frac{dU}{\sqrt{U-U_T}},
\end{equation}
so
\begin{equation}
U-U_T=\frac{3U_T^{1/2}}{4R^{3/2}}r^2=\frac{3U_T^{1/2}}{4R^{3/2}}(X^2+Y^2).
\label{k}
\end{equation}
Thus
\begin{equation}
\frac{ds^2}{\alpha'}\approx \frac{9U_T}{4R^3} r^2dt^2+dr^2.
\end{equation}
We have 
\begin{equation}
R^{3/2}=g_5\sqrt{N/4\pi}=g_{YM}\sqrt{N/4\pi T}.
\end{equation}
Because of periodicity of the angle, i.e. identification of 
$\theta\rightarrow \theta+2\pi$, corresponding to $t\rightarrow
t+1/T$ (i.e. $T=b/\pi$), one therefore needs
\begin{equation}
\theta^2=\frac{9U_T}{4R^3}t^2,~{\rm i.e.}~T=\frac{3U_T^{1/2}}
{2g_5\sqrt{\pi N}}.
\end{equation}
Using (\ref{k}) we then have
\begin{equation}
U=U_T+\frac{3U_T^{1/2}}{4R^{3/2}}(X^2+Y^2).
\end{equation}

We can now proceed as in the 3-d case. The expanded action is 
\begin{equation}
S\approx \frac{1}{2\pi}\int d^2\sigma \left(\frac{U^{3/2}}{R^{3/2}}+
\frac{1}{2}\left[(\partial_i X)^2+(\partial_i Y)^2\right]\right).
\end{equation}
Using
\begin{equation}
\frac{U^{3/2}}{R^{3/2}}\approx\frac{U_T^{3/2}}{R^{3/2}}+\frac{9U_T}{8R^3}
(X^2+Y^2),
\end{equation}
this leads to an $X,Y$ (former $Z,T$) -dependent integrand
\begin{equation}
\frac{1}{2}\left[(\partial_i X)^2+(\partial_i Y)^2+\frac{9U_T}{4R^3}
(X^2+Y^2)\right].
\label{mass}
\end{equation}
with mass parameter
\begin{equation}
\frac{9U_T}{4R^3}=4\pi^2T^2.
\end{equation}
We can now compute the contribution to the potential using the results in ref
\cite{sungkil}, and adding the leading 
terms (ignoring terms which are exponentially
small), we get the string tension in four dimensions by use of analytic
regularization ($\ln x=\pa x^\beta/\pa\beta$ for $\beta\rightarrow 0$)
\bea
{\cal T}_4 &=&\frac{8\pi}{27}g_{YM}^2NT^2
\left[1+\frac{27}{2g_{YM}^2N}\left(1
-\ln\frac{4\pi^2T^2}{\mu^2}\right)\right]
\non \\
           &=&\frac{8\pi}{27}g_{YM}^2NT^2
\left[1-\frac{27}{2g_{YM}^2N}\ln\frac{4\pi^2T^2}{e\mu^2}\right].
\label{4dstring}
\eea
Here $\mu$ is the arbitrary scale introduced in the last section.

We end this section by remarking again that the effective flatness of
AdS space, in the strong-coupling limit, suggests that the 
fermi and ghost degrees of freedom contribute to the L\"uscher term
as in flat space.  If this is so, then the naive counting argument
of the last section again leads to the net result for the L\"uscher term
\beq
+{\pi\over 12L}.
\label{result}
\eeq
in the quark-antiquark potential. 
This should be compared to what has been used in fits to the lattice
Monte Carlo data, namely
\beq
-{(d-2)\pi\over 24L}=-{\pi\over 12L}.
\label{res}
\eeq
Thus the magnitude is the same, but the signs are opposite.  

This result is based on the assumption that the worldsheet fermions 
essentially live in a flat space, so that the flat string action is relevant. 
Although we
have given some plausibility arguments, this remains to be proven. However,
if the result is true, it has been pointed out to us by L\"uscher \cite{L}  
that this has the far reaching consequence that supergravity
in the limit $g^2_{YM}N\rightarrow\infty$ has nothing to do with QCD defined
on the lattice, for any $N$ and with or without matter fields. The reason is 
that it has been shown rigorously by Bachas \cite{Bachas} that the heavy 
quark potential $V(L)$ is monotonic and concave. Thus, 
\beq
V'(L)\geq 0~{\rm and}~V''(L)\leq 0
\eeq
for all $L$, contradicting a positive value of the L\"uscher term.

\setcounter{equation}{0} 
\section{Massive fields and extrinsic curvature}

   One of the most interesting questions in non-perturbative gauge 
theory, which the AdS/CFT correspondence may eventually address, concerns 
the form of the effective D=4 string theory describing the QCD string.
In this connection, we would like to make a remark that may be relevant
for the understanding of the existence of massive fields versus
reparametrization invariance.

   When the 1-loop contributions of two massive and two massless worldsheet 
modes are combined, one finds a result which is strongly reminiscent
of string models with extrinsic curvature \cite{extrinsic}. 
The extrinsic curvature $K^i_{ab}$ (=the second fundamental form) is given by
\beq
D_aD_b {\bf X}=K^i_{ab}{\bf N}_i,~{\rm with}~{\bf N}_i{\bf N}_k=
\delta_{ik}~{\rm and}~{\bf N}_i\pa_a{\bf X}=0,
\eeq
where ${\bf X}(\sigma,\tau)$ is is the position vector for some surface, 
${\bf N}_i$ are the normals, and $D_a$ is the covariant
derivative with respect to the induced metric $g_{ab}=\pa_a X\pa_bX$. There 
are many expressions for the extrinsic curvature. Here we need in particular
\beq
K^{ia}_bK^{ib}_a=\left(\pa_a(\sqrt{g}g^{ab}\pa_bX)\right)^2.
\eeq
Thus the extrinsic curvature is of fourth order in the derivatives.

    It was noticed in ref. 
\cite{sungkil} that a perturbative expansion of the string with
extrinsic curvature leads to tr ln's coming from massive fields. This can
be seen by use of the relation
\bea
{\rm tr}\ln\left(-\nabla^2\right)+{\rm tr}\ln\left(-\nabla^2+4\pi^2T^2\right)
&=&{\rm tr}\ln\left(\frac{1}{2}\mu_0(-\nabla^2)+\frac{\lambda_T}{27\pi}
(-\nabla^2)^2\right)
\non \\
  & & -{\rm tr}\ln\frac{\lambda_T}{27\pi},
\label{add}
\eea
where $\mu_0=8\pi\lambda_TT^2/27$ is the string tension to leading order,
and $\lambda_T$ is the 't Hooft coupling $g^2_{YM}N$. 
The left hand side of 
this equation combines the Gaussian integrations over four world sheet fields:
The first term on the
left hand side can be taken from two of the
massless string fields, whereas the second term comes from the two
massive fields. The combined tr ln on the right hand side can be
considered as coming from the effective action\footnote{The last term
in (\ref{add}) can be absorbed in the constant $\mu$, which is
anyhow arbitrary: $\mu^2\rightarrow 27\pi\mu^2/\lambda_T$.}
\beq
S_{\rm eff}=\int d^2\sigma\left[\mu_0+\frac{1}{2}\mu_0(\pa_aX)^2+
\frac{\lambda_T}{27\pi}(\pa_a^2X)^2\right],
\label{1q}
\eeq
where we added the leading term $\mu_0~YT$.
However, this effective action can in turn be considered \cite{sungkil}  as 
the perturbative version of
\beq
S_{\rm eff}=\int d^2\sigma\left[\mu_0\sqrt{g}+\frac{\lambda_T}
{27\pi}\sqrt{g}K^{ia}_bK^{ib}_a\right],
\label{2}
\eeq
where (\ref{1q}) arises from (\ref{2}) by a perturbative expansion
of the metric and the determinant by use of
\beq
g_{ab}=\delta_{ab}+\pa_a X\pa_b X,
\eeq
keeping only terms of order $X^2$. The $X$'s here are 2 dimensional and 
transverse. In (\ref{2}) there are four $X$'s, two of which are longitudinal,
so we are looking at a four-dimensional theory of extrinsic curvature, and 
an effective string of positive rigidity.\footnote{In contrast, vortex tubes
found in abelian Higgs models appear to have negative rigidity, and may
be unstable at the quantum level \cite{vortex}.}


For a superstring in flat space with, e.g., Ramond boundary conditions, 
the bosonic tr ln's exactly cancel the 
fermionic ones. In our case, we have argued that the fermions still
live in an effectively flat space. Hence the total result of the
Gaussian integrations is
\beq
-{\rm tr}\ln [-\nabla^2]+{\rm tr}\ln [-\nabla^2+4\pi^2T^2]=
-2{\rm tr}\ln [-\nabla^2]+{\rm tr}\ln [(-\nabla^2)^2+4\pi^2T^2(-\nabla^2)].
\label{FF}
\eeq
The last term on the right hand side has the interpretation in
terms of extrinsic curvature discussed above, and can be formulated as
in eq. (\ref{2}). The first term on the right hand side of (\ref{FF}) can 
be considered as the contribution from fermions,
\beq
S_F={\rm const}\int d^2\sigma\left(\psi
\partial\llap{/}\psi\right).
\eeq
Here $\psi (\sigma,\tau)$ is a two-dimensional 
Majorana spinor which is also a four
dimensional vector, and $S_F$ should be added to $S_{\rm eff}$ in
eq. (\ref{2}). Also, the boundary conditions on $\psi$ are
that they should be of the Ramond type, i. e. $\psi_1 (-L/2,\tau)=
\psi_2(-L/2,\tau)$ and $\psi_1(L/2,\tau)=\psi_2(L/2,\tau)$.

Thus, in D=4 dimensions, we can view the trace log contributions
as arising from an effective four dimensional string theory, which has 
both extrinsic curvature and worldsheet fermions.
It is of interest that in the ``effective''
picture one does not see all the extra dimensions (although 
these may show up at higher orders in $1/g_{YM}^2N$). It should be noted,
however, that this effective theory, as it stands, is associated with a 
``wrong-sign'' (i.e. repulsive) L\"uscher term.

\setcounter{equation}{0}
\section{Conclusions}

   We have found that two of the bosonic modes of the Maldacena-Witten 
worldsheet are massive.  These mass terms are relevant for the 
existence of a L{\"u}scher term in the heavy quark potential, and they
may also be related to extrinsic curvature terms in an effective $D=4$
string theory. Concerning the L\"uscher term, our very tentative 
conclusion is that such a term
appears, and in four dimensions it has the same magnitude, but opposite
sign of the one used in fits to lattice Monte Carlo data. The basis for
this result is the discussion in Section 3, according to which the
eigenvalues of worldsheet Laplacians are essentially like those in 
flat space, and we also expect flat space contributions from the fermions 
and ghosts.  In flat space, whether we consider
Ramond boundary conditions, or (more relevant to the AdS case) 
Neveu-Schwartz boundary conditions with the tachyon removed by the 
GSO projection, the L\"uscher term vanishes.  If two bosonic modes
become massive, they do not contribute to the L\"uscher term, and a
naive counting argument suggests that the net L\"uscher term has the
wrong sign.  This argument does not, however, take into account the
Ramond-Ramond background, and there may of course be surprises 
encountered when the full boson-fermion action in the black hole AdS 
background becomes known.

In the absence of such surprises in the fermion and/or ghost sectors,
it would appear that the heavy quark potential extracted from the 
AdS/CFT correspondence at $g^2_{YM} N \ra \infty$ is in 
qualitative disagreement with lattice QCD at any coupling.  
This is not only a matter of
disagreement with the trend in the Monte Carlo data, where fits are subject to
uncertainties. It was pointed out by L\"uscher \cite{L} that Bachas has 
proven, on the basis of reflection positivity alone, that
the heavy quark potential in lattice QCD is both monotonic increasing and
convave \cite{Bachas}, and that this implies an attractive or vanishing, 
rather than repulsive, $1/L$ term. Hence there is no point in
trying to fit Monte Carlo data to a positive $1/L$ coefficient before
the weak coupling limit $g^2_{YM}N\rightarrow 0$ is reached.  The
AdS/CFT approach is a very different regulator from lattice gauge
theory.  If the concavity of the potential is violated at strong couplings,
then at these couplings the AdS/CFT solution cannot adequately
represent the long-range physics of continuum QCD.

  Another difference between AdS/CFT and lattice QCD at strong-couplings,
which we have previously noted \cite{Us}, is that 
the glueball mass remains finite while the string tension diverges
as $g^2_{YM} N \ra \infty$, whereas in lattice QCD the string tension 
and glueball mass spectrum both diverge in this limit.  

   It is clearly very important to see if worldsheet fermion and/or ghost
contributions could somehow change our conclusions, at least in regard
to the sign of the L\"uscher term, and for this purpose it will be necessary
to investigate the full action of the fermionic sector in the AdS black-hole 
background.  Curvature is negligible and the saddlepoint configuration
is essentially constant (except near the 
$\s \ra \pm L/2$ endpoints) as $R,L \ra \infty$, which suggests but
by no means proves that flat-space fermion/ghost contributions would
be obtained in those limits.  It is possible, when the full fermionic
action (together with the Ramond-Ramond background) is taken into account,
that a L\"uscher term in agreement with lattice Monte Carlo may be
obtained.  But it is also possible that qualitative agreement 
between the AdS/CFT and lattice formulations of planar gauge theory can 
only be obtained away from the strong-coupling limit, perhaps after a 
phase transition as discussed in ref.\ \cite{Gross}.  

\vspace{33pt}

\ni {\Large \bf Acknowledgements}

\bigskip

   We have benefited from
discussions with Korkut Bardakci, Martin L\"uscher, Hirosi Ooguri, and 
Peter Orland. J.G.'s research is supported by the U.S.\ Department of Energy
under Grant.\ No.\ DE-FG03-92ER40711.

\end{document}